\documentclass[pra,floatfix,reprint,amsmath,amssymb,aps]{revtex4-2}

\usepackage{graphicx}
\usepackage{bm}
\usepackage{dcolumn}
\usepackage{placeins}
\usepackage{url}
\usepackage{subfig}
\usepackage{booktabs}
\usepackage{tikz}
\usetikzlibrary{arrows.meta,decorations.pathmorphing}
\usepackage{enumitem}

\begin{document}

\preprint{APS/123-QED}

\title{Clarifying the energy storage capabilities of virtual critical coupling\\ under realistic excitation constraints}

\author{Valentin Mazières$^1$, Théo Delage$^2$, Jérôme Sokoloff$^3$, Olivier Pascal$^3$}
\affiliation{$^1$ISAE-SUPAERO, Université de Toulouse, 31055 Toulouse, France\\ $^2$ENAC, Université de Toulouse, Toulouse, France\\ $^3$LAPLACE, Université de Toulouse, CNRS, INPT, UPS, Toulouse, France}

\begin{abstract}
Virtual critical coupling (VCC) has emerged as a promising approach for achieving reflectionless excitation of resonant systems through tailored incident waveforms. However, the energy storage enhancement often attributed to VCC is generally assessed without considering practical constraints imposed by the excitation source. In this work, we revisit the energy storage capabilities of VCC under a realistic maximum-amplitude constraint. Using temporal coupled-mode theory, we derive analytical expressions for the stored energy of continuous wave (CW), ideal VCC (IVCC), and constrained VCC (CVCC) excitations. While the conventional IVCC excitation leads to higher stored energy than CW excitation due to its exponentially increasing incident amplitude, we show that this enhancement originates from the larger incident energy delivered by the unconstrained waveform. When the maximum excitation amplitude is fixed, the proposed CVCC excitation stores less energy than CW excitation throughout the excitation duration, while preserving the high energy transfer efficiency and reflectionless excitation enabled by VCC. The analytical expressions are used to analyze both an ideal lossless resonator and an experimentally lossy microwave cavity previously used for plasma ignition by VCC. The results clarify that VCC should primarily be regarded as a method for improving energy transfer efficiency and controlling transient excitation, rather than as an intrinsic mechanism for increasing the absolute stored energy under realistic source limitations.
\end{abstract}

\maketitle

\section{Introduction}

Efficiently transferring electromagnetic energy into resonant systems is a fundamental objective in many areas of physics and engineering, including microwave engineering, photonics, and plasma generation. Conventional resonator excitation relies on monochromatic continuous wave (CW) signals tuned to the resonance frequency. Under these conditions, the maximum energy that can be stored inside the resonator is determined by the balance between the incident power, the external coupling, and the intrinsic losses. In particular, for overcoupled resonators, a significant fraction of the incident power is reflected, limiting the efficiency of the energy transfer.

Virtual perfect absorption, also known as virtual critical coupling (VCC) in one-port resonators, has recently emerged as an alternative excitation strategy capable of eliminating reflections during the transient excitation of an overcoupled resonator by employing incident signals with a complex excitation frequency \cite{baranov_coherent_2017,radi_virtual_2020,trainiti_coherent_2019,zhong_coherent_2020,delage_experimental_2022,araujo-martinez_virtual_2025}. Under these conditions, the incident waveform grows exponentially in time, leading to a reflectionless transient excitation and an efficient transfer of the incident energy into the resonator. These properties have motivated considerable interest in VCC for applications such as wave control, photonics \cite{krasnok_anomalies_2019,kim_complex-frequency_2025}, and, more recently, microwave plasma ignition \cite{delage_plasma_2023}.

The performance of VCC has often been assessed by comparing the stored energy obtained under VCC excitation with that obtained under a conventional CW excitation \cite{radi_virtual_2020}. Such comparisons have often been interpreted as suggesting that VCC leads to an increase in the stored energy within the resonator. However, the exponentially growing incident signal used in the theoretical description of VCC reaches amplitudes that can largely exceed those of the CW excitation. Consequently, the two excitations are generally not compared under identical source constraints. Since practical waveform generators are limited by a maximum available output amplitude, comparing CW and VCC excitations without accounting for this limitation may lead to misleading conclusions regarding their practical energy storage capabilities.

The objective of the present work is therefore to revisit the comparison between CW and VCC excitations under a realistic maximum-amplitude constraint. To this end, we distinguish the conventional exponentially growing excitation, referred to here as ideal virtual critical coupling (IVCC), from a constrained virtual critical coupling (CVCC) excitation obtained by rescaling the IVCC waveform so that its maximum amplitude is identical to that of the CW excitation. Analytical expressions for the stored energy are first derived using temporal coupled-mode theory, allowing a direct comparison between CW, IVCC, and CVCC excitations. These analytical expressions are then used to analyze both an ideal lossless cavity and an experimental lossy microwave cavity previously used for plasma ignition. Finally, the implications of these results for the interpretation of VCC performance reported in the literature are discussed.

\section{Problem formulation}

\subsection{Theoretical framework}

\begin{figure*}[t]
\centering

\begin{tikzpicture}[
    >=Latex,
    thick,
    scale=1.15
]


\draw[blue, thick] (-7.5,1.88) -- (-3.5,1.88);
\node[blue] at (-4.05,2.2) {$|\underline{s_+^{CW}}(t)|$};

\draw[green!50!black, thick, domain=0:1, samples=100, variable=\x]
plot (
{-7.5+4.0*\x},
{1.88+2.1*(exp(4*\x)-1)/(exp(4)-1)}
);
\node[green!50!black] at (-5.1,3) {$|\underline{s_+^{IVCC}}(t)|$};

\draw[red, thick, domain=0:1, samples=100, variable=\x]
plot (
{-7.5+4.0*\x},
{-0.2+2.1*(exp(4*\x)-1)/(exp(4)-1)}
);
\node[red] at (-5.1,1) {$|\underline{s_+^{CVCC}}(t)|$};

\draw[->,thick] (-7.5,-0.3) -- (-7.5,4.1);
\node[left] at (-7.5,1.8) {$s_{max}$};
\node[anchor=east, align=right] at (-7.7,3.6) {\small Incident signal\\amplitude};

\draw[->,thick] (-8,-0.2) -- (-3.1,-0.2);
\draw[thick] (-3.5,-0.3) -- (-3.5,-0.1);
\node at (-7.5,-0.55) {$0$};
\node at (-3.5,-0.55) {$t_f$};
\node at (-2.95,-0.2) {$t$};


\draw[->,thick] (-2,1.5) -- (-1.2,1.5);

\draw[fill=gray!15] (0,1.5) circle (1.2cm);
\node at (0,1.8) {$\underline{a}(t)$};
\node at (0,1.35) {$\omega_0$};

\draw[->,orange,very thick,
    decorate,
    decoration={
        snake,
        amplitude=1mm,
        segment length=3mm
    }
] (0.85,2.3) -- (1.9,3.1);

\node[orange] at (2.3,3.2) {$\gamma_{int}$};

\draw[<->,magenta,very thick,dashed]
  (-1.8,1.7) to[out=60,in=120] (-0.7,1.7);
\node[magenta] at (-1.3,2.25) {$\gamma_{ext}$};

\draw[->,thick]
(-1.2,1.5)
.. controls (-1.2,1.1) and (-1.5,0.9) ..
(-2.0,0.9);

\node[left] at (-2.0,0.9) {$\underline{s_-}(t)$};
\node[left] at (-2.0,1.5) {$\underline{s_+}(t)$};

\end{tikzpicture}

\caption{
Schematic representation of the resonator excitation considered in this work.
The same resonator is excited by different incident signals:
continuous wave (CW), ideal virtual critical coupling (IVCC), and constrained virtual critical coupling (CVCC).
}

\label{fig:theoretical_background}
\end{figure*}
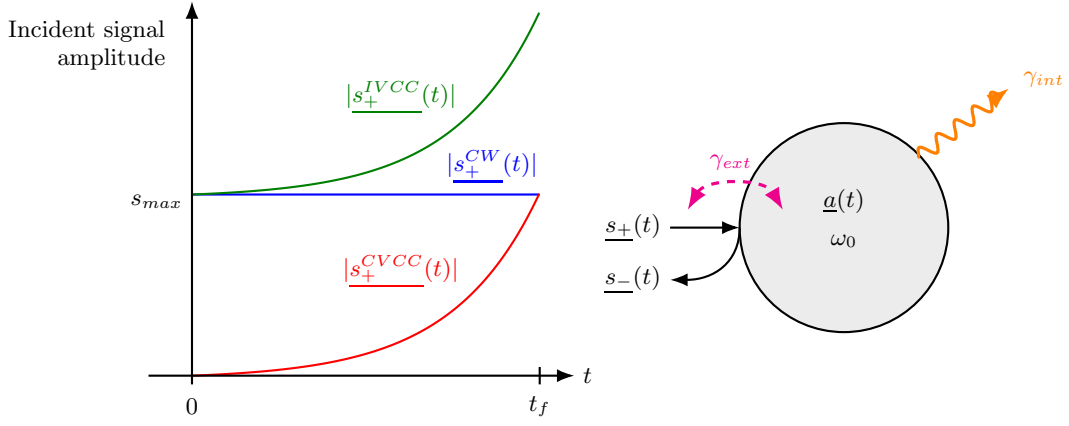

The dynamics of a single resonant mode is described using temporal coupled-mode theory (an $e^{-j\underline{\omega}t}$ time convention is assumed) \cite{haus_waves_1984,suh_temporal_2004}:

\begin{equation}
\frac{d\underline{a}(t)}{dt}
=
(-j\omega_0-\gamma)\underline{a}(t)
+\sqrt{2\gamma_{ext}}\,\underline{s_+}(t),
\label{eq_a_t}
\end{equation}

where $\underline{a}(t)$ is the complex modal amplitude, $\omega_0$ is the resonance frequency, $\gamma=\gamma_{ext}+\gamma_{int}$ is the total decay rate, and $\underline{s_+}(t)$ is the incident signal. The outgoing signal is given by

\begin{equation}
\underline{s_-}(t)
=
-\underline{s_+}(t)
+\sqrt{2\gamma_{ext}}\,\underline{a}(t)
\label{eq_s_moins}
\end{equation}

where $\underline{s_-}(t)$ is the reflected signal from the resonator, as illustrated in figure \ref{fig:theoretical_background}.

Solving equation (\ref{eq_a_t}) for an initially uncharged resonator ($\underline{a}(0)=0$) driven by the incident signal

\begin{align}
\underline{s_+}(t)
&=
A_+e^{-j\underline{\omega}t}
=
A_+e^{-j\omega_0t}e^{\omega''t}
\end{align}

where $A_+$ is a real constant and $\underline{\omega}=\omega_0+j\omega''$, yields

\begin{equation} 
\underline{a}(t)=\frac{\sqrt{2\gamma_{ext}}A_+}{\gamma+\omega''}e^{-j\omega_0 t}\left(e^{\omega'' t}-e^{-\gamma t}\right)
\end{equation}

This gives the energy stored in the cavity mode:

\begin{equation}
U(t)=|\underline{a}(t)|^2=\frac{2\gamma_{ext}A_+^2}{(\gamma+\omega'')^2}\left(e^{\omega'' t}-e^{-\gamma t}\right)^2\label{eq_U_general}
\end{equation}

Thus, the temporal evolution of the energy stored in the resonator depends on the resonator parameters ($\gamma_{ext}$ and $\gamma_{int}$), the incident signal amplitude $A_+$, and its exponential growth rate $\omega''$. This expression provides the basis for the following analysis, where the performance of virtual critical coupling (VCC) is compared with that of continuous wave (CW) excitation.

\subsection{Excitation methods}

The different excitation signals associated with the methods considered in this work are represented in figure \ref{fig:theoretical_background}, and are listed below:

\begin{itemize}[leftmargin=10pt]
    \item The continuous wave (CW) excitation corresponds to the classical harmonic signal used to excite resonators:
\begin{align}
\underline{s_+^{CW}}(t)=s_{max}e^{-j\omega_0t}\label{eq_s_inc_CW}
\end{align}
with $A_+^{CW}=s_{max}$ and $\omega_{CW}=\omega_0$.
\item The ideal virtual critical coupling (IVCC) excitation corresponds to the conventional description of VCC, with the associated incident signal \cite{radi_virtual_2020,baranov_coherent_2017}:

\begin{align}
\underline{s_+^{IVCC}}(t)=s_{max}e^{-j\omega_0t}e^{\omega''_{VCC}t}\label{eq_s_inc_IVCC}
\end{align}
with $A_+^{IVCC}=s_{max}$ and $\underline{\omega}_{VCC}=\omega_0+j\omega_{VCC}''$, where $\omega_{VCC}''=\gamma_{ext}-\gamma_{int}$ \cite{radi_virtual_2020,baranov_coherent_2017}. It corresponds to an exponentially growing signal that starts at $s_{max}$ at $t=0$, as represented on figure \ref{fig:theoretical_background}. In the following, the term IVCC refers to the ideal exponential excitation commonly used in theoretical studies of VCC \cite{radi_virtual_2020}. Note that in reference \cite{radi_virtual_2020}, the incident signals are multiplied by a smoothing function so that their initial amplitudes are close to zero. This function rapidly approaches unity over time, so that the expressions in equations (\ref{eq_s_inc_CW}) and (\ref{eq_s_inc_IVCC}) are recovered (with an additional factor of 0.1 for the VCC excitation case \cite{radi_virtual_2020}, as discussed below).
\item The constrained virtual critical coupling (CVCC) excitation, introduced in the present work:
\begin{align}
\underline{s_+^{CVCC}}(t)=
\frac{s_{max}}{e^{\omega_{VCC}''t_f}}
e^{-j\omega_0t}e^{\omega''_{VCC}t}\label{eq_s_inc_CVCC}
\end{align}
with $A_+^{CVCC}=s_{max}/e^{\omega_{VCC}''t_f}$. It corresponds to an exponentially increasing signal terminating at $s_{max}$ at $t=t_f$, where $t_f$ is the duration of the incident signal, as shown in figure \ref{fig:theoretical_background}.
\end{itemize}

Table \ref{tab:parameters} summarizes the definitions of the parameters used throughout the analysis presented in this paper. Throughout this work, we consider only overcoupled cavities, \textit{i.e.}, cavities satisfying $\gamma_{ext}>\gamma_{int}$. Therefore, all conclusions drawn in this paper apply exclusively to this regime, unless otherwise specified. 

\begin{table}[h]
\centering
\renewcommand{\arraystretch}{1.5}
\begin{tabular}{rcl@{\hspace{1.5cm}}rcl}
\toprule
\toprule
$\gamma$                    & $=$ & $\gamma_{ext}+\gamma_{int}$ &
$Q_{ext}$                   & $=$ & \(\omega_0/(2\gamma_{ext})\) \\

$\underline{\omega}$        & $=$ & $\omega_0+j\omega''$ &
$Q_{int}$                   & $=$ & \(\omega_0/(2\gamma_{int})\) \\

$\omega_{CW}$               & $=$ & $\omega_0$ &
$A_{+}^{CW}$                & $=$ & $s_{max}$ \\

$\underline{\omega}_{VCC}$  & $=$ & $\omega_{0}+j\omega_{VCC}''$ &
$A_{+}^{IVCC}$              & $=$ & $s_{max}$ \\

$\omega_{VCC}''$            & $=$ & $\gamma_{ext}-\gamma_{int}$ &
$A_{+}^{CVCC}$              & $=$ & $s_{max}/e^{\omega_{VCC}''t_f}$ \\

\bottomrule
\bottomrule
\end{tabular}
\caption{Parameters used in the analysis.}
\label{tab:parameters}
\renewcommand{\arraystretch}{1}
\end{table}

\subsection{Motivation}

CW and IVCC excitations form the basis of the analysis performed on the energy storage capabilities of these methods in resonators \cite{radi_virtual_2020}. However, from a practical point of view, the maximum amplitude that can be generated by the waveform generator is limited to a given value $s_{max}$. If the maximum amplitude of the CW signal is set to this value, a fair comparison between CW and VCC excitations requires the maximum amplitude of the VCC signal to be limited to the same value. This limitation can significantly affect the comparison of the energy storage performances of these methods and may lead to misleading conclusions regarding their practical performance. Addressing this issue is the objective of the present paper. To this end, we introduce the CVCC excitation, whose incident signal is obtained by rescaling the amplitude of the IVCC signal to satisfy the maximum available amplitude constraint $s_{max}$.

\section{Energy Storage Capabilities of VCC Methods Relative to CW}

We are interested in this section in finding the expressions of the energy ratio, defined as the ratio of the cavity stored energy obtained with a VCC method to that obtained with a CW excitation.

To this end, first the expression of the CW stored energy is deduced from equation (\ref{eq_U_general}) and the corresponding expression of the incident signal (equation (\ref{eq_s_inc_CW})):

\begin{align}
U_{CW}(t)=\frac{2\gamma_{ext}s_{max}^2}{\gamma^2}\left(1-e^{-\gamma t}\right)^2\label{eq_U_CW}
\end{align}    


\subsection{IVCC}

The energy stored in the cavity mode under IVCC excitation is obtained from equations (\ref{eq_U_general}) and (\ref{eq_s_inc_IVCC}):

\begin{align}
U_{IVCC}(t)=\frac{s_{max}^2}{2\gamma_{ext}}\left(e^{\omega_{VCC}'' t}-e^{-\gamma t}\right)^2\label{eq_U_IVCC}
\end{align} 

The corresponding IVCC energy factor, defined as the ratio between the stored energy obtained with IVCC and CW excitations, is therefore:

\begin{align}
F_{IVCC}(t)&=\frac{U_{IVCC}(t)}{U_{CW}(t)}
=
\frac{1}{4}\left(
\frac{\gamma}{\gamma_{ext}}
\frac{e^{\omega_{VCC}'' t}-e^{-\gamma t}}
{1-e^{-\gamma t}}
\right)^2\label{eq_F_IVCC}
\end{align}

As demonstrated in appendix \ref{append_F_IVCC}, for an overcoupled cavity this factor is always superior to 1 for $t\leq t_f$. This relation $F_{IVCC}(t)>1$ leads to:

\begin{equation}
U_{\mathrm{IVCC}}(t)>U_{CW}(t),
\qquad 0<t\leq t_f 
\end{equation}

Hence, for the conventional IVCC excitation considered here, where the initial amplitude is identical to that of the CW signal, the stored energy in the cavity mode is higher than for CW excitation. This result is not surprising, as the IVCC incident signal has a larger amplitude than the CW signal during the excitation process (as illustrated in  figure \ref{fig:theoretical_background}) and is specifically designed to suppress the reflected signal from the resonator.

\subsection{CVCC}

The energy stored in the cavity mode under CVCC excitation is obtained from equations (\ref{eq_U_general}) and (\ref{eq_s_inc_CVCC}):

\begin{align}
U_{CVCC}(t)=e^{-2\omega_{VCC}'' t_f}\frac{s_{max}^2}{2\gamma_{ext}}\left(e^{\omega_{VCC}'' t}-e^{-\gamma t}\right)^2\label{eq_U_CVCC}
\end{align} 

The corresponding CVCC energy factor, defined as the ratio between the stored energy obtained with CVCC and CW excitations, is therefore:

\begin{align}
F_{CVCC}(t)&=\frac{U_{CVCC}(t)}{U_{CW}(t)}\nonumber\\
&=
\frac{e^{-2\omega''_{VCC} t_f}}{4}\left(
\frac{\gamma}{\gamma_{ext}}
\frac{e^{\omega''_{VCC} t}-e^{-\gamma t}}
{1-e^{-\gamma t}}
\right)^2\label{eq_F_CVCC}
\end{align}

As demonstrated in appendix \ref{append_F_CVCC}, for an overcoupled cavity this factor is always inferior to 1 for $t\leq t_f$. This relation $F_{CVCC}(t)<1$ leads to:

\begin{equation}
U_{\mathrm{CVCC}}(t)<U_{CW}(t)\label{eq_U_CVCC_inf_CW},
\qquad 0<t\leq t_f
\end{equation}

Therefore, under the same maximum excitation amplitude constraint, throughout the excitation duration, CVCC results in lower stored energy in the cavity mode than CW excitation. This remains true even though reflection is present in the CW case, whereas it is suppressed with CVCC. Although the VCC process enables higher energy efficiency (up to unity in the lossless case \cite{radi_virtual_2020}), the absolute stored energy remains lower than that obtained with CW excitation when both excitations are subjected to the same maximum amplitude constraint. This difference arises from the amplitude normalization applied to the exponentially increasing VCC waveform, which reduces the incident energy delivered to the cavity.

\begin{table}[h]
\centering
\renewcommand{\arraystretch}{1.2}
\setlength{\tabcolsep}{10pt} 
\label{tab:enhancement_factor}
\begin{tabular}{lc}
\hline
\hline
\textbf{Overcoupled cavity} ($\gamma_{ext}>\gamma_{int}$) & $F_{CVCC}^{max}$ \\
\hline
~ & ~\\
Lossy cavity ($\gamma_{int}\ne0$) &
$\displaystyle
\frac{1}{4}\frac{(\gamma_{ext}+\gamma_{int})^2}{\gamma_{ext}^2}$ \\

~ & ~\\

Lossless cavity ($\gamma_{int}=0$) &
$\displaystyle
\frac{1}{4}$ \\

~ & ~\\
\hline
\hline
\end{tabular}
\caption{Maximum CVCC energy factor for the different cavity configurations.}
\label{tab:enhancement_factor}
\renewcommand{\arraystretch}{1}
\end{table}

The maximum value of the factor $F_{CVCC}^{max}$ is reached at the end $t_f$ of the incident signal. Assuming that the cavity mode has reached its quasi-steady-state, denoted by the superscript $ss$ ($F^{ss}_{CVCC}(t)=F_{CVCC}(t\gg1/\tau)$), equation (\ref{eq_F_CVCC}) becomes at $t=t_f$:

\begin{align} 
F_{CVCC}^{max}=F_{CVCC}^{ss}(t_f)&= \frac{1}{4}\left( \frac{\gamma_{ext}+\gamma_{int}}{\gamma_{ext}} \right)^2\label{eq_F_CVCC_t_f}
\end{align}

The maximum CVCC energy factor is summarized in Table~\ref{tab:enhancement_factor} for different cavity configurations.

\section{Validations}

In this section, we validate the theoretical predictions derived in the previous section. To this end, two cavities are considered: the theoretical lossless cavity studied in reference \cite{radi_virtual_2020} and the experimental lossy cavity used in reference \cite{delage_plasma_2023} for plasma ignition using VCC. The first case enables us to validate our theoretical predictions by comparing our IVCC results with those reported in reference \cite{radi_virtual_2020} and to interpret the behavior of the proposed CVCC excitation in the lossless case. The second experimental case enables us to interpret the behavior of the proposed CVCC excitation in the lossy case and provides insight into the advantages and limitations of using VCC instead of CW for plasma ignition.

To this end, as in reference \cite{radi_virtual_2020}, two additional quantities are also evaluated:

\begin{itemize}
    \item The efficiency, defined as
    \begin{equation}
    \eta(t)=
    \frac{|\underline{a}(t)|^2}
    {\int_0^t |\underline{s_+}(\tau)|^2\,d\tau}\label{eq_eta}
    \end{equation}

    \item The reflection coefficient, defined as
    \begin{equation}
    r(t)=
    \left|\frac{\underline{s_-}(t)}
    {\underline{s_+}(t)}\right|^2
    \end{equation}
\end{itemize}

\subsection{Theoretical lossless cavity}

\begin{figure}[h!]
    \centering

    \subfloat[CW and IVCC excitations.]{
        \includegraphics[width=\columnwidth]{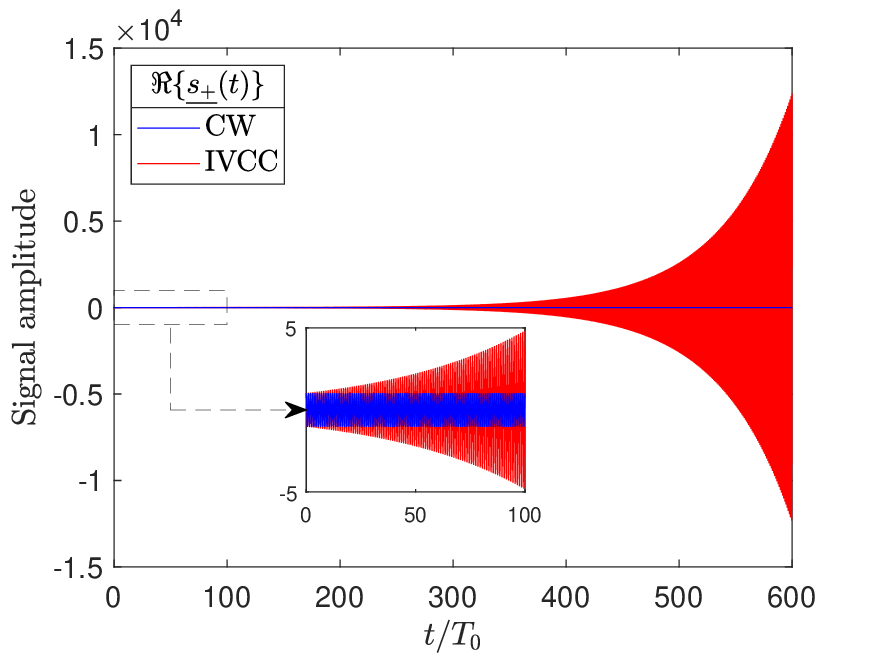}
        \label{fig:s_IVCC}
    }

    \vspace{0.3cm}

    \subfloat[CW and CVCC excitations.]{
        \includegraphics[width=\columnwidth]{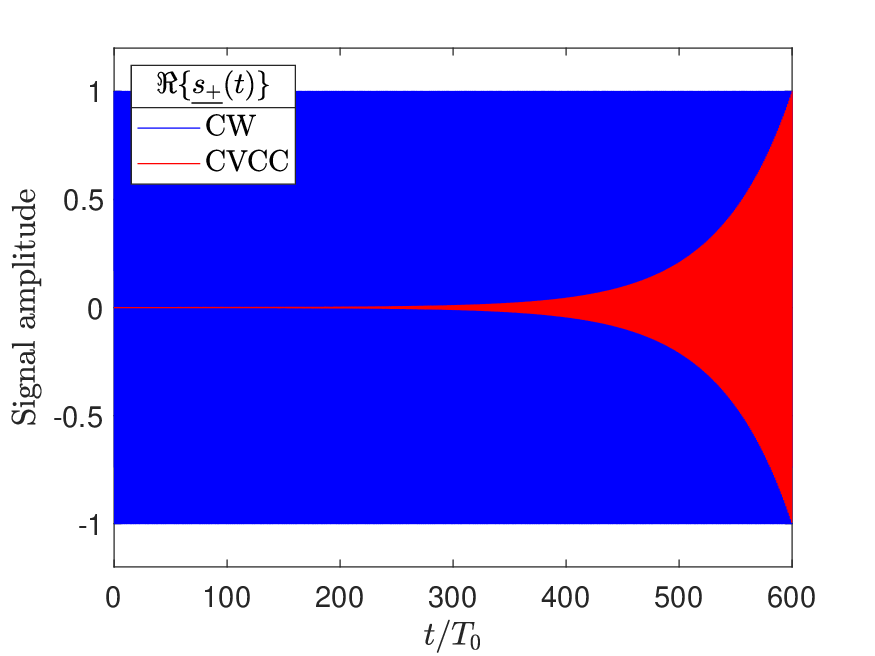}
        \label{fig:s_CVCC}
    }

    \caption{
    Incident signals used for CW, IVCC and CVCC excitations of the theoretical lossless cavity of reference \cite{radi_virtual_2020}.
    }
    \label{fig:s_IVCC_CVCC_krasnok}
\end{figure}

\begin{figure*}[t]
\centering

\subfloat[]{
\includegraphics[width=0.45\textwidth]{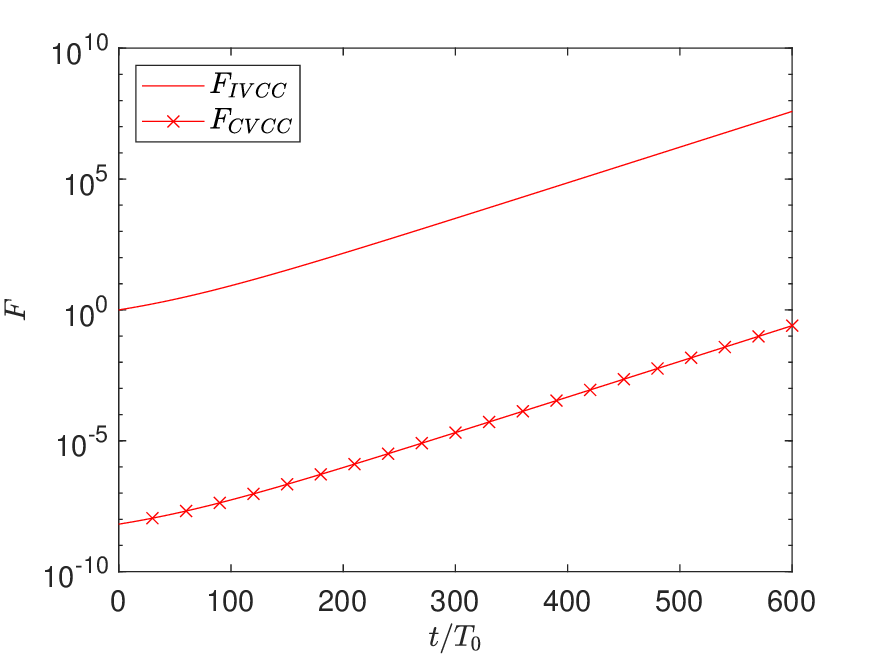}
\label{fig:a}
}
\hfill
\subfloat[]{
\includegraphics[width=0.45\textwidth]{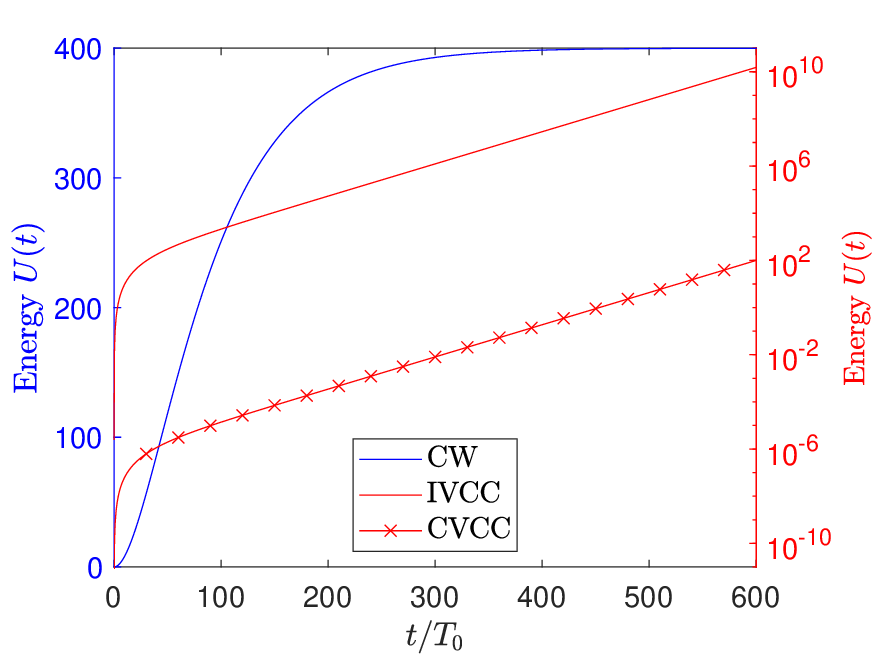}
\label{fig:b}
}

\vspace{0.3cm}

\subfloat[]{
\includegraphics[width=0.45\textwidth]{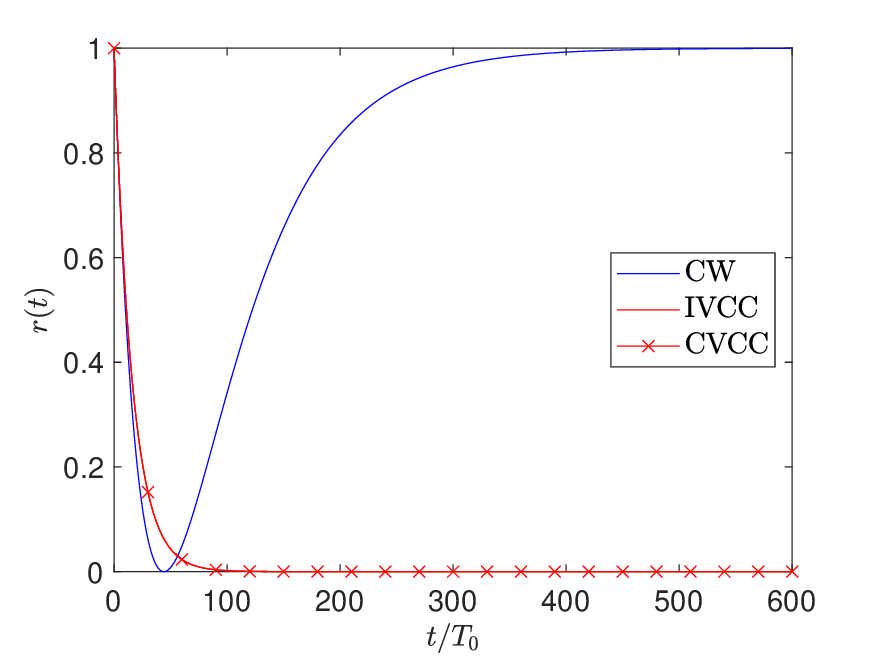}
\label{fig:c}
}
\hfill
\subfloat[]{
\includegraphics[width=0.45\textwidth]{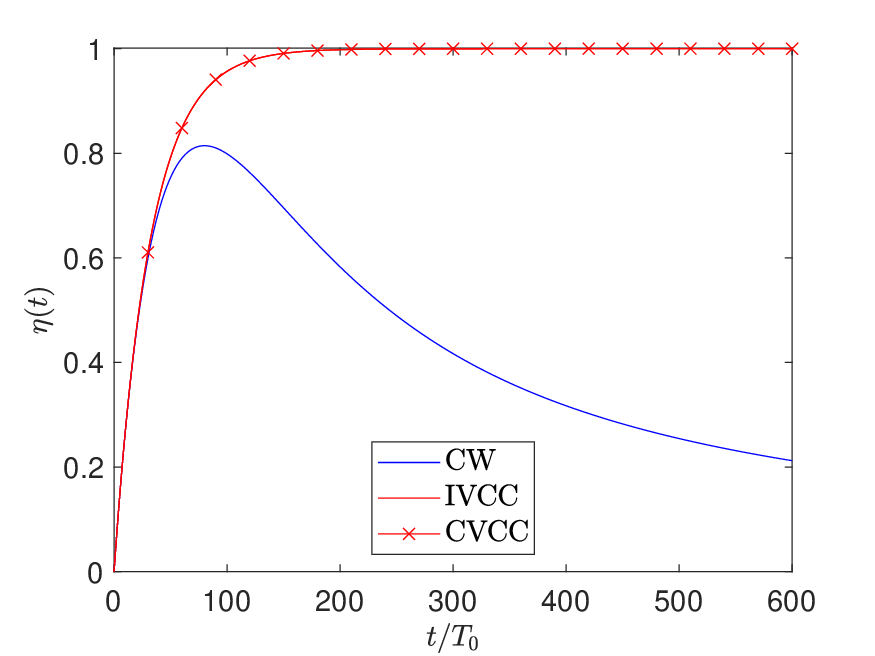}
\label{fig:d}
}

\caption{
Comparison between CW, IVCC and CVCC excitations of the theoretical lossless cavity of reference \cite{radi_virtual_2020}.
(a) Energy factor.
(b) Stored energy in the cavity mode.
(c) Reflection coefficient.
(d) Efficiency.
}
\end{figure*}

We consider as an example the theoretical lossless cavity studied in reference \cite{radi_virtual_2020}. The corresponding parameters are shown in table \ref{tab:parameters_krasnok}. The incident signals are constructed using equations (\ref{eq_s_inc_CW}), (\ref{eq_s_inc_IVCC}), and (\ref{eq_s_inc_CVCC}), and plotted in figure \ref{fig:s_IVCC_CVCC_krasnok} (with $s_{max}=1$). 

\begin{table}[h]
\centering
\renewcommand{\arraystretch}{1.5}
\begin{tabular}{rcl@{\hspace{1.5cm}}rcl}
\toprule
\toprule
$\omega_0$        & $=$ & $2$      &
$Q_{\rm ext}$     & $=$ & $200$ \\

$T_0$             & $=$ & $\pi$    &
$\gamma_{\rm ext}$& $=$ & $0.005$  \\

$t_f/T_0$         & $=$ & $600$   &
$Q_{\rm int}$     & $=$ & $\infty$ \\

          &   &      &
 $\gamma_{\rm int}$& $=$ & $0$      \\
\bottomrule
\bottomrule
\end{tabular}
\caption{Parameters of the theoretical lossless cavity of reference \cite{radi_virtual_2020}.}
\label{tab:parameters_krasnok}
\renewcommand{\arraystretch}{1}
\end{table}

As expected, the IVCC incident signal is significantly larger than the incident CW signal, reaching a value of about 12400 at \(t=t_f\), compared with about 1 for the latter. The CVCC incident signal reaches the maximum amplitude of the CW incident signal at \(t=t_f\), in agreement with the expected behavior. Both the CW and CVCC signals can be generated using the same waveform generator system, which is the basis of the CVCC concept. We will now examine the consequences of this amplitude constraint on the resulting performances.

The energy factor, stored energy in the cavity mode, reflection coefficient, and efficiency of the three excitations (CW, IVCC, and CVCC) are shown in figures  \ref{fig:a}, \ref{fig:b}, \ref{fig:c}, and \ref{fig:d}, respectively.

For the CW excitation, the stored energy shown in figure \ref{fig:b} increases until reaching its steady-state value, with a characteristic timescale of \(1/\gamma\), as described by equation (\ref{eq_U_CW}). In the present lossless case, this value is \(2/\gamma_{\mathrm{ext}}=400\), obtained from the same equation with \(\gamma_{\mathrm{int}}=0\) and \(s_{max}=1\). The associated reflection coefficient, shown in figure \ref{fig:c}, starts at 1, as expected from equation (\ref{eq_s_moins}) with \(a(0)=0\), decreases to zero, and then gradually increases back to 1, as expected for an overcoupled cavity \cite{radi_virtual_2020,delage_experimental_2022}. Finally, the efficiency, shown in figure \ref{fig:d}, increases until reaching a maximum of about 0.8, and then decreases to a value close to 0.2. This behavior is consistent with the definition of the efficiency given in equation (\ref{eq_eta}): as the time integrated incident energy in the denominator continues to increase, while the stored energy in the numerator saturates, the efficiency tends to zero.

For the IVCC excitation, the stored energy increases exponentially in time, as expected from equation (\ref{eq_U_IVCC}) in the quasi-steady-state regime ($t\gg1/\gamma$), and in agreement with the results reported in reference \cite{radi_virtual_2020}. As expected from the developments presented in this paper, the energy factor remains greater than 1 for all times, starting from unity at $t=0$. The reflection coefficient rapidly decreases to zero, while the efficiency increases toward unity, in agreement with the expected behavior \cite{radi_virtual_2020}.

The results shown in figures \ref{fig:b}, \ref{fig:c}, and \ref{fig:d} for the CW and IVCC excitations are in good agreement with those reported in reference \cite{radi_virtual_2020}. Note that slight differences are observed in the transient regime and in the final value of the stored energy for the VCC excitation. These differences arise from the fact that, in reference \cite{radi_virtual_2020}, a smoothing function is applied to the incident signals, together with an additional factor of 0.1 for the VCC signal. This additional scaling explains why the final stored energy reported in reference \cite{radi_virtual_2020} differs by a factor of 100 from that obtained for the IVCC excitation in figure \ref{fig:b}. Nevertheless, the overall behavior remains the same, as does the physical interpretation.

For the CVCC excitation, the reflection coefficient and the efficiency are identical to those obtained with the IVCC excitation, as expected. However, the stored energy in the cavity mode is significantly lower with CVCC because of the lower amplitude of the incident signal (see figure \ref{fig:s_IVCC_CVCC_krasnok}). As expected from the developments presented in this paper, the energy factor remains lower than 1 for all times. Hence, compared with the CW excitation, the stored energy also remains much lower. Furthermore, at the end of the excitation (\(t=t_f\)), the stored energy is four times lower with CVCC than that obtained with the CW excitation, as predicted by equation (\ref{eq_F_CVCC_t_f}) in the lossless case and reported in table \ref{tab:enhancement_factor}. This is the case even though a strong reflection is present with CW excitation ($r(t)$ close to one as time increases) whereas the reflection is equal to zero with CVCC. This highlights that minimizing the reflected power alone does not necessarily maximize the energy stored in the cavity.

\subsection{Experimental lossy cavity}

\begin{figure}[h!]
    \centering

    \subfloat[CW and IVCC excitations.]{
        \includegraphics[width=\columnwidth]{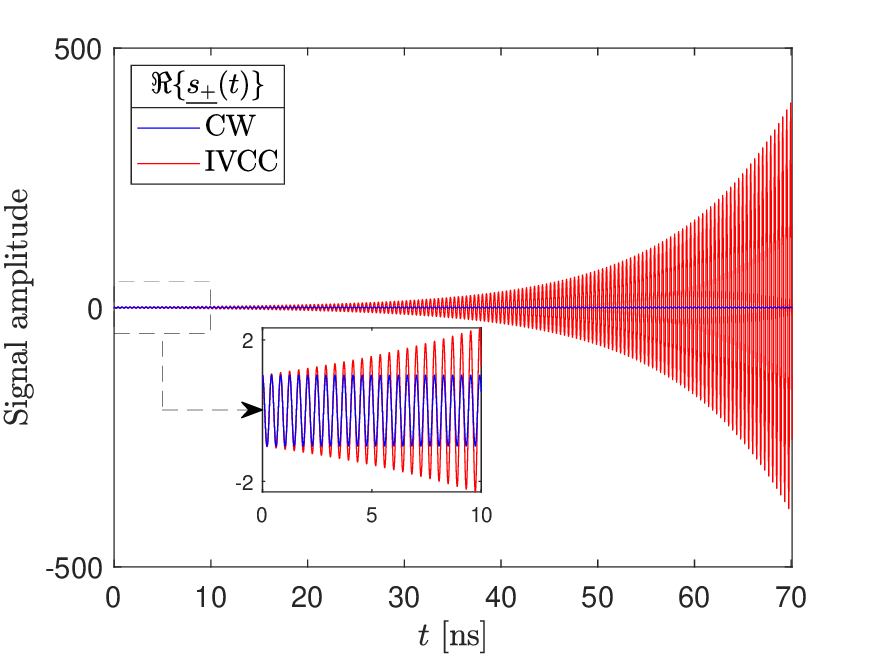}
        \label{fig:s_Theo_IVCC}
    }

    \vspace{0.3cm}

    \subfloat[CW and CVCC excitations.]{
        \includegraphics[width=\columnwidth]{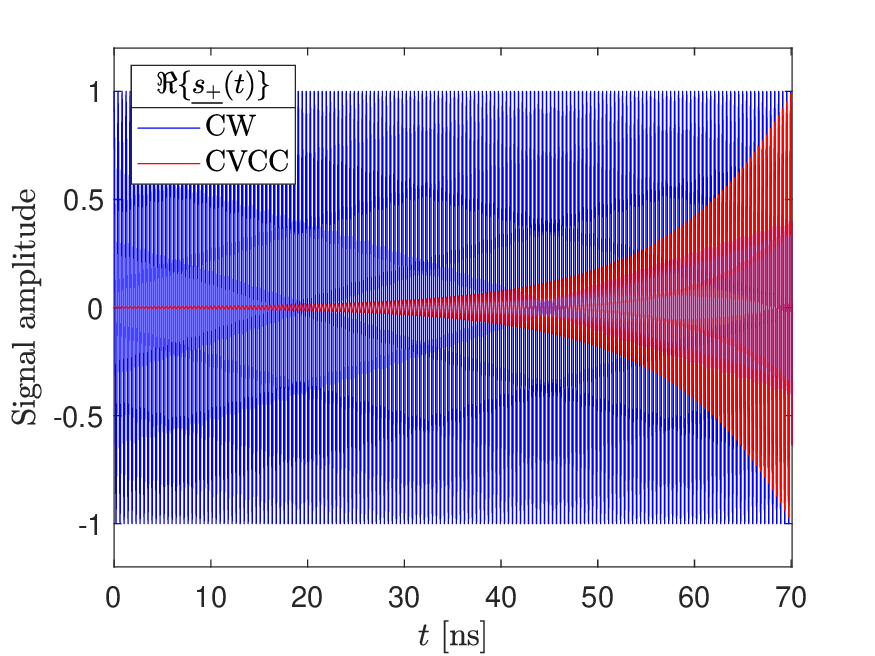}
        \label{fig:s_Theo_CVCC}
    }

    \caption{
    Incident signals used for CW, IVCC, and CVCC excitations of the experimental lossy cavity of reference \cite{delage_plasma_2023}.
    }
    \label{fig:s_IVCC_CVCC_Theo}
\end{figure}

We consider now the experimental lossy cavity studied in reference \cite{delage_plasma_2023}. The corresponding parameters are shown in table \ref{tab:parameters_Theo}. The incident signals are constructed using equations (\ref{eq_s_inc_CW}), (\ref{eq_s_inc_IVCC}), and (\ref{eq_s_inc_CVCC}), and plotted in figure \ref{fig:s_IVCC_CVCC_Theo} (with $s_{max}=1$). 

\begin{table}[h]
\centering
\renewcommand{\arraystretch}{1.5}
\begin{tabular}{rcl@{\hspace{1.5cm}}rcl}
\toprule
\toprule
$\omega_0$         & $=$ & $2\pi\times2.416~\mathrm{GHz}$ &
$Q_{\rm ext}$      & $=$ & $84$ \\

$T_0$              & $=$ & $0.41~\mathrm{ns}$ &
$Q_{\rm int}$      & $=$ & $1417$ \\

$t_f$              & $=$ & $70~\mathrm{ns}$ &
$\gamma_{\rm ext}$ & $=$ & $90.810\times10^6~\mathrm{s^{-1}}$ \\

       &   &    &
$\gamma_{\rm int}$ & $=$ & $5.356\times10^6~\mathrm{s^{-1}}$ \\

\bottomrule
\bottomrule
\end{tabular}
\caption{Parameters of the experimental lossy cavity of reference \cite{delage_plasma_2023}.}
\label{tab:parameters_Theo}
\renewcommand{\arraystretch}{1}
\end{table}

\begin{figure*}[t]
\centering

\subfloat[]{
\includegraphics[width=0.45\textwidth]{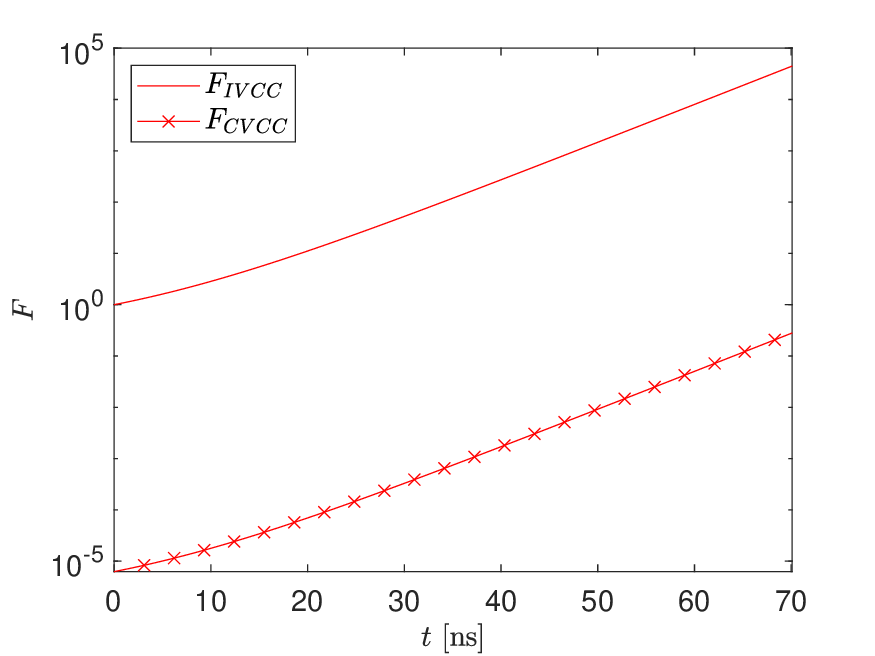}

\label{fig:a_Theo}
}
\hfill
\subfloat[]{
\includegraphics[width=0.45\textwidth]{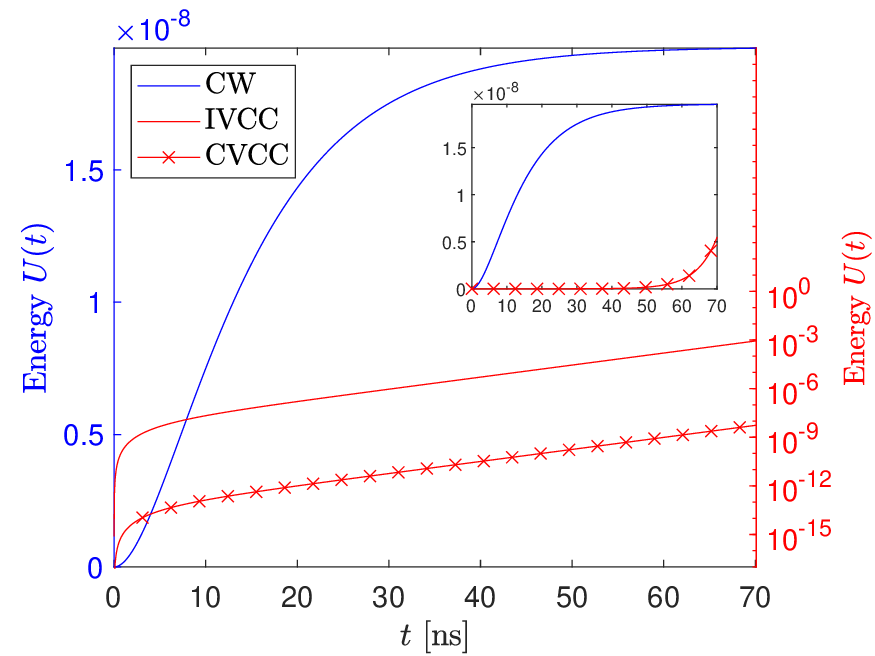}
\label{fig:b_Theo}
}

\vspace{0.3cm}

\subfloat[]{
\includegraphics[width=0.45\textwidth]{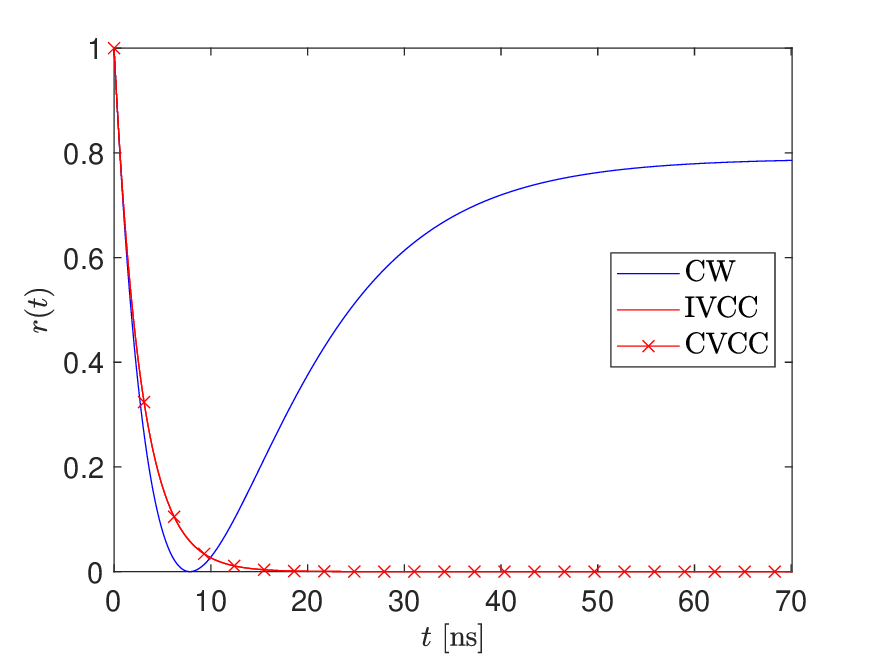}
\label{fig:c_Theo}
}
\hfill
\subfloat[]{
\includegraphics[width=0.45\textwidth]{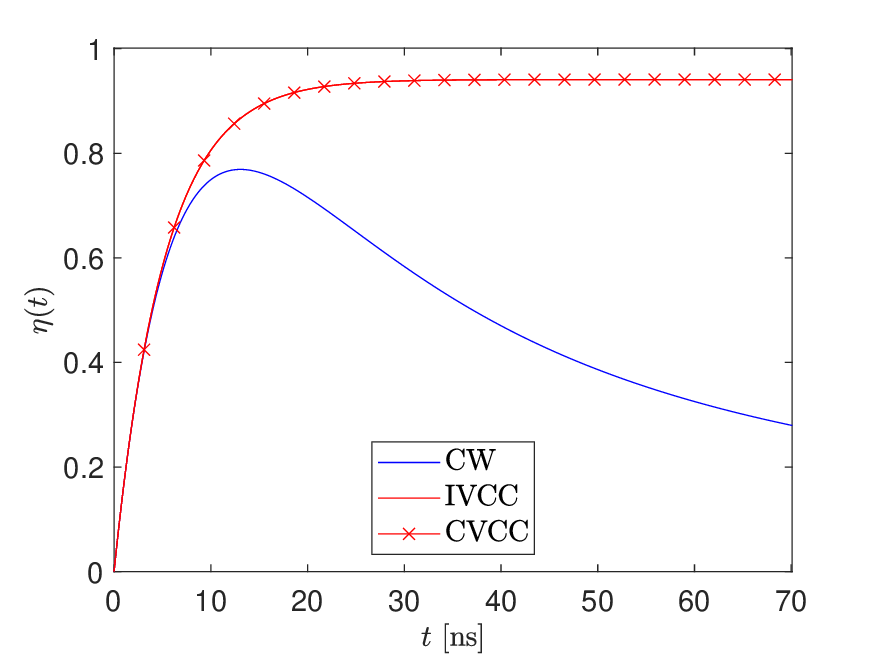}
\label{fig:d_Theo}
}

\caption{
Comparison between CW, IVCC and CVCC excitations of the experimental lossy cavity of reference \cite{delage_plasma_2023}.
(a) Energy factor.
(b) Stored energy in the cavity mode. Inset: CW and CVCC plotted on the same linear scale.
(c) Reflection coefficient.
(d) Efficiency.
}

\end{figure*}

In this case, the IVCC incident signal is larger than the incident CW signal, reaching a value of about 400 at \(t=t_f\), compared with about 1 for the latter. The CVCC incident signal reaches the maximum amplitude of the CW incident signal at \(t=t_f\), in agreement with the expected behavior. Both the CW and CVCC signals can be generated using the waveform generator system employed to ignite plasmas in reference \cite{delage_plasma_2023}. Although \cite{delage_plasma_2023} reports results obtained with VCC excitation, it does not compare them with those obtained using CW excitation. The following comparison highlights the advantages and limitations of VCC relative to CW excitation, particularly for plasma ignition applications, as discussed in the next section.

The energy factor, stored energy in the cavity mode, reflection coefficient, and efficiency of the three excitations (CW, IVCC, and CVCC) are shown in figures \ref{fig:a_Theo}, \ref{fig:b_Theo}, \ref{fig:c_Theo}, and \ref{fig:d_Theo}, respectively.

The trends obtained with this experimental lossy cavity are similar to those obtained with the theoretical lossless cavity because both cavities are overcoupled. The main differences are that the reflection coefficient for the CW excitation does not return to 1 and the efficiency of the VCC methods does not reach unity, owing to the intrinsic losses of the experimental cavity. Compared with the CW excitation, the stored energy with CVCC also remains much lower, in agreement with our theoretical predictions (see equation (\ref{eq_U_CVCC_inf_CW})). Furthermore, at the end of the excitation (\(t=t_f\)), the stored energy with CVCC is approximately one third of that obtained with CW excitation, as predicted by equation (\ref{eq_F_CVCC_t_f}) for the lossy case and reported in table \ref{tab:enhancement_factor} (using the parameters reported in table \ref{tab:parameters_Theo}). This difference is illustrated in the inset of figure \ref{fig:b_Theo}. This remains true even though a strong reflection is present with the CW excitation, with \(r(t)\) approaching 0.8 as time increases, whereas the reflection coefficient remains zero with the CVCC excitation. This highlights that, even with a lossy cavity, minimizing the reflected power alone does not necessarily maximize the energy stored in the cavity.

\section{Clarifying the interpretation of VCC performance}

Virtual critical coupling (VCC) has been widely recognized for its ability to suppress reflections and to maximize the transfer of the incident energy into the cavity mode \cite{radi_virtual_2020,baranov_coherent_2017}. However, this improvement should not be interpreted as an enhancement of the absolute energy stored in the cavity. Under a fixed maximum excitation amplitude, the amount of energy that can be delivered to the cavity is itself limited. As demonstrated in the present work, although VCC provides nearly reflectionless excitation, a CW excitation stores more energy than the corresponding amplitude-constrained VCC excitation (CVCC). VCC should therefore be regarded primarily as a technique for improving energy transfer efficiency rather than for maximizing the absolute stored energy. Clarifying this distinction helps avoid possible misinterpretations of the practical performance of VCC. In the following, we revisit several results reported in the literature in light of the findings of the present work.

\subsection{Normalized versus absolute stored energy}

It has previously been reported that VCC yields an eightfold enhancement of the intracavity intensity compared with monochromatic excitation at the real valued resonance frequency \cite{hinney_efficient_2024}. Such comparisons rely on normalized quantities that characterize the efficiency of the energy transfer from the source to the cavity. They do not address the absolute stored energy when the excitation is constrained by the maximum available source amplitude. The present work shows that these two viewpoints lead to fundamentally different conclusions. Specifically, under the same maximum-amplitude constraint, CVCC stores less energy than CW and therefore produces a lower intracavity field.

\subsection{Interpretation of the factor of four}

Recent works have also interpreted a factor of four predicted by coupled-mode theory as an intrinsic enhancement of the stored energy achieved with VCC compared with CW excitation in the critical coupling regime \cite{akram_transiently_2026}. This interpretation, however, does not correspond to the physical origin of this factor. This factor of four, originally reported in reference \cite{radi_virtual_2020}, arises from the comparison of the steady-state stored energies of two different resonators driven by the same monochromatic CW excitation: an overcoupled lossless resonator ($\gamma_{ext}\gg\gamma_{int}$) and a critically coupled resonator ($\gamma_{ext}=\gamma_{int}$). In the following, the subscripts $OC$ and $CC$ refer to overcoupled and critically coupled resonators, respectively, while superscript $ss$ denotes the steady-state or quasi-steady-state value of the stored energy ($U^{ss}(t)=U(t\gg1/\tau)$). Using equation (\ref{eq_U_CW}), one directly obtains

\begin{equation}
U^{\mathrm{ss}}_{\mathrm{CW,CC}}
=
\frac{U^{\mathrm{ss}}_{\mathrm{CW,OC}}}{4}
=\frac{s_{max}^2}{2\gamma_{ext}}
 \label{eq_U_ss}
\end{equation}

This factor is therefore a consequence of the different coupling conditions and does not correspond to an enhancement generated by a VCC excitation.

By contrast, the comparison addressed in the present work concerns the same overcoupled resonator excited either by CW or by VCC. Under the same maximum-amplitude constraint, the CVCC quasi-steady-state expression obtained from equation (\ref{eq_U_CVCC}), along with (\ref{eq_U_ss}) give

\begin{align}
U^{\mathrm{ss}}_{\mathrm{CVCC,OC}}(t)
=\frac{s_{max}^2}{2\gamma_{ext}}e^{2\omega_{VCC}'' (t-t_f)}
\le
U^{\mathrm{ss}}_{\mathrm{CW,CC}}
=
\frac{U^{\mathrm{ss}}_{\mathrm{CW,OC}}}{4}
\end{align}

where the equality is reached only at the end of the excitation duration, at $t=t_f$:

\begin{align}
U^{\mathrm{ss}}_{\mathrm{CVCC,OC}}(t_f)
=
U^{\mathrm{ss}}_{\mathrm{CW,CC}}
=
\frac{U^{\mathrm{ss}}_{\mathrm{CW,OC}}}{4}
\end{align}

Therefore, the factor of four discussed in reference \cite{akram_transiently_2026} does not represent an intrinsic enhancement of the stored energy produced by VCC. Under a fixed maximum-amplitude constraint, CVCC in an overcoupled resonator leads to the same stored energy as CW excitation in a critically coupled resonator, but four times less energy than CW excitation in the same overcoupled resonator.

\subsection{Implications for plasma ignition}

VCC has recently been proposed for microwave plasma ignition \cite{delage_plasma_2023}. Its main advantage is the suppression of reflections during the transient excitation, allowing an efficient transfer of the incident microwave energy into the resonator. In addition, the temporal profile of the excitation could be tailored to the dynamics of plasma formation, providing additional flexibility for optimizing the ignition process. However, under a fixed maximum source amplitude, the present results show that a CW excitation produces a larger stored energy, and therefore a larger intracavity electric field than the corresponding amplitude-constrained VCC excitation. Consequently, the practical interest of VCC for plasma ignition lies primarily in its reflectionless excitation and waveform engineering capabilities rather than in an increase of the maximum electromagnetic energy inside the cavity, as suggested in reference \cite{akram_transiently_2026}.

\section{Conclusion}

In this work, we investigated the energy storage capabilities of virtual critical coupling under a realistic maximum excitation amplitude constraint. To distinguish the conventional theoretical excitation from its experimentally realizable counterpart, we introduced the concept of constrained virtual critical coupling (CVCC), obtained by normalizing the amplitude of the ideal VCC excitation to the same maximum value as the reference CW excitation.

Analytical expressions derived from temporal coupled-mode theory demonstrate that, for overcoupled resonators, the ideal VCC excitation always stores more energy than a CW excitation because its exponentially growing incident waveform reaches significantly larger amplitudes. In contrast, when both excitations are subjected to the same maximum source amplitude, the opposite conclusion is obtained: the proposed CVCC excitation always stores less energy than the corresponding CW excitation throughout the excitation process. This result was confirmed for both an ideal lossless cavity and an experimentally lossy microwave cavity.

These results clarify the physical interpretation of VCC. The principal advantage of VCC is not an intrinsic enhancement of the absolute stored energy, but rather its ability to suppress reflections and maximize the fraction of the incident energy transferred into the resonator. Consequently, metrics based on normalized energy transfer efficiency should not be directly interpreted as enhancements of the absolute stored energy under realistic source constraints.

Beyond clarifying the interpretation of previous VCC studies, the present work provides a practical framework for assessing the performance of VCC in applications where the excitation source is limited by a maximum available amplitude. In particular, for microwave plasma ignition, the results show that the practical interest of VCC lies primarily in its reflectionless excitation and waveform engineering capabilities rather than in an increase of the maximum electromagnetic field achievable inside the resonator.

\newpage
\appendix

\section{Demonstration of $F_{IVCC}(t)>1$}\label{append_F_IVCC}

To prove that $F_{IVCC}(t)>1$, we first rewrite this condition as (knowing that $\omega_{VCC}''>0$ and assuming $t>0$):

\begin{equation}
\frac{\gamma}{\gamma_{ext}}
\frac{e^{\omega''_{VCC}t}-e^{-\gamma t}}
{1-e^{-\gamma t}}>2 
\end{equation}

Because $\gamma>0\rightarrow1-e^{-\gamma t}>0$ this inequality can be rewritten as

\begin{equation}
\gamma
e^{\omega''_{VCC}t}
+
\omega''_{VCC}
e^{-\gamma t}
-2\gamma_{ext}>0 
\end{equation}

We introduce the function $g(t)$ as:

\begin{equation}
g(t)=\gamma
e^{\omega''_{VCC}t}
+
\omega''_{VCC}
e^{-\gamma t}
-2\gamma_{ext}
\end{equation}

At $t=0$,

\begin{equation}
g(0)=0 .
\end{equation}

Its derivative is

\begin{align}
g'(t)
&=
\gamma\omega''_{VCC}\left(e^{\omega''_{VCC}t}-e^{-\gamma t} \right)
\end{align}

Since the cavity is overcoupled

\begin{equation}
\omega''_{VCC}>0
\end{equation}

and therefore

\begin{equation}
g'(t)>0
\qquad (t>0)
\end{equation}

Hence, $g(t)$ is strictly increasing for $t>0$ and thus:

\begin{equation}
g(t)>g(0)=0
\end{equation}

which proves that

\begin{equation}
\boxed{
F_{IVCC}(t)>1,
\qquad t>0 
}
\end{equation}

~

\section{Demonstration of $F_{CVCC}(t)<1$}\label{append_F_CVCC}

To prove that $F_{CVCC}(t)<1$, we first rewrite this condition as

\begin{equation}
F_{\mathrm{CVCC}}(t)
=
e^{2\omega''_{VCC}(t-t_f)}
\left(
\frac{\gamma}{2\gamma_{ext}}
\frac{1-e^{-2\gamma_{ext}t}}
{1-e^{-\gamma t}}
\right)^2 
\end{equation}

We now introduce the function

\begin{equation}
h(x)=\frac{1-e^{-xt}}{x}
\end{equation}

which is monotonically decreasing with respect to $x$ for $x>0$. For an overcoupled cavity:

\begin{equation}
\gamma_{ext}>\gamma_{int}\rightarrow 2\gamma_{ext}>\gamma
\end{equation}

which gives

\begin{equation}
h(2\gamma_{ext})
<
h(\gamma)
\end{equation}

This leads to

\begin{equation}
\frac{\gamma}{2\gamma_{ext}}
\frac{1-e^{-2\gamma_{ext}t}}
{1-e^{-\gamma t}}<1
\end{equation}

and thus

\begin{equation}
0<\left(\frac{\gamma}{2\gamma_{ext}}
\frac{1-e^{-2\gamma_{ext}t}}
{1-e^{-\gamma t}}\right)^2<1
\end{equation}

Moreover, since $t\leq t_f$ and $\omega''_{VCC}>0$,

\begin{equation}
0<e^{2\omega''_{VCC}(t-t_f)}\leq1
\end{equation}

Consequently,

\begin{equation}
\boxed{
F_{\mathrm{CVCC}}(t)<1,
\qquad 0<t\leq t_f 
}
\end{equation}

\bibliographystyle{unsrt}
\bibliography{Article_VCC}

\end{document}